\documentclass[10pt, twocolumn]{IEEEtran}
\usepackage{amsmath, mathrsfs,amsfonts}
\usepackage{amsfonts}
\usepackage{amssymb}
\usepackage{acronym}
\usepackage{graphicx}

\def\ignore#1{}

\def\bs{\boldsymbol}

\newtheorem{assumption}{Assumption}

\newtheorem{dom_face}{Definition}

\newtheorem{optimal_policy}[dom_face]{Definition}
\newtheorem{greedy_policy}[dom_face]{Definition}
\newtheorem{approximate_policy}[dom_face]{Definition}
\newtheorem{improved_approximate_policy}[dom_face]{Definition}

\newtheorem{expansion}[dom_face]{Definition}
\newtheorem{Hausdorff}[dom_face]{Definition}
\newtheorem{approx_proj}[dom_face]{Definition}

\newtheorem{convergence-thm}{Theorem}
\newtheorem{tracking_worst}[convergence-thm]{Theorem}

\newtheorem{iteration_queue}[convergence-thm]{Theorem}
\newtheorem{tracking_avg}[convergence-thm]{Theorem}

\newtheorem{rate-splitting}[approx_proj_prop]{Proposition}

\newtheorem{half-space-proj}{Lemma}
\newtheorem{psuedo-nonexp}[half-space-proj]{Lemma}

\newtheorem{opt_dist}[half-space-proj]{Lemma}
\newtheorem{region_dist}[half-space-proj]{Lemma}
\newtheorem{subgradient_rate}[half-space-proj]{Lemma}

               {\begin{list}{}{\leftmargin#1\rightmargin#2\topsep#3}\item{}}%
               {\end{list}}

\begin{document}
\title{Dynamic Rate Allocation in Fading Multiple Access Channels\thanks{This research was partially supported by the
National Science Foundation under grant DMI-0545910, and by DARPA
ITMANET program.}}


\author{Ali ParandehGheibi\thanks{A.\ ParandehGheibi is with the Laboratory for
Information and Decision Systems, Electrical Engineering and
Computer Science Department, Massachusetts Institute of Technology,
Cambridge MA, 02139 (e-mail: parandeh@mit.edu)}, Atilla
Eryilmaz\thanks{A.\ Eryilmaz is with the Electrical and Computer
Engineering, Ohio State University, OH, 43210 (e-mail:
eryilmaz@ece.osu.edu)}, Asuman Ozdaglar, and Muriel M\'edard\thanks{
A.\ Ozdaglar and M.\ M\'edard are with the Laboratory for
Information and Decision Systems, Electrical Engineering and
Computer Science Department, Massachusetts Institute of Technology,
Cambridge MA, 02139 (e-mails: asuman@mit.edu, medard@mit.edu)}}


\maketitle \thispagestyle{headings}

\begin{abstract}
We consider the problem of rate allocation in a fading Gaussian
multiple-access channel (MAC) with fixed transmission powers. Our
goal is to maximize a general concave utility function of
transmission rates over the throughput capacity region. In contrast
to earlier works in this context that propose solutions where a
potentially complex optimization problem must be solved in every
decision instant, we propose a low-complexity approximate rate
allocation policy and analyze the effect of temporal channel
variations on its utility performance. To the best of our knowledge,
this is the first work that studies the tracking capabilities of an
approximate rate allocation scheme under fading channel conditions.

We build on an earlier work to present a new rate allocation policy
for a fading MAC that implements a low-complexity approximate
gradient projection iteration for each channel measurement, and
explicitly characterize the effect of the speed of temporal channel
variations on the tracking neighborhood of our policy. We further
improve our results by proposing an alternative rate allocation
policy for which tighter bounds on the size of the tracking
neighborhood are derived. These proposed rate allocation policies
are computationally efficient in our setting since they implement a
single gradient projection iteration per channel measurement and
each such iteration relies on approximate projections which has
polynomial-complexity in the number of users.
\end{abstract}

\section{Introduction}
Dynamic allocation of communication resources such as bandwidth or
transmission power is a central issue in multiple access channels in
view of the time-varying nature of the channel and interference
effects. Most of the existing literature on resource allocation in
multiple access channels focuses on specific communication schemes
such as TDMA (time-division multiple access) \cite{TDMA} and CDMA
(code-division multiple access) \cite{CDMA1,CDMA3} systems. An
exception is the work by Tse \emph{et al.} \cite{Tse}, who
introduced the notion of \emph{throughput capacity} for the fading
channel with Channel State Information (CSI) and studied dynamic
rate allocation policies with the goal of maximizing a linear
utility function of rates over the throughput capacity region.

An important literature relevant to our work appears in the context
of cross-layer design, where joint scheduling-routing-flow control
algorithms have been proposed and shown to achieve utility
maximization for concave utility functions while guaranteeing
network stability (e.g. \cite{linshr05, erysri05, neemodli05,
sto05}). The common idea behind these schemes is to use properly
maintained queues to make dynamic decisions about new packet
generation as well as rate allocation.

Some of these works (\cite{erysri05, neemodli05}) explicitly address
the fading channel conditions, and show that their policies can
achieve rates arbitrarily close to the optimal based on a design
parameter choice. However, the rate allocation imposed by these
schemes requires that a large optimization problem requiring global
information be solved over a complex feasible set in every time
slot. Clearly, this may not always be possible due to the
limitations of the available information, or the processing power,
or the complexity intrinsic to the feasible set. In fact, even in
the absence of fading, the interference constraints between nearby
nodes' transmissions may make the feasible set so complex that the
optimal rate allocation problem becomes NP-hard (see
\cite{eryozdmod07}).

In the absence of fading, several works have proposed and analyzed
approximate randomized and/or distributed rate allocation algorithms
for various interference models (\cite{tas98, linshr05, modshazus06,
eryozdmod07, sanbuisri07, joolinshr07}), and their effect on the
utility maximization is investigated in
\cite{eryozdmod07,eryozdshamod07}. However, no similar work exists
for fading channel conditions, where the changes in the fading
conditions coupled with the inability to solve the optimization
problem instantaneously make the solution much more challenging. In
fact, it is not even clear what algorithm can be used to achieve
close to close-to-optimal performance.

In this work, we propose an approximate gradient projection method
and study its tracking capabilities when the channel conditions vary
over time. In our algorithm, the solution is updated in every time
slot in a direction to increase the utility function at that time
slot. But, since the channel may vary between time-slots, the extend
of these temporal channel variations become critical to the
performance. We explicitly quantify the impact of the speed of
fading on the performance of the policy, both for the worst-case and
the average behavior. Our results also capture the effect of the
degree of concavity of the utility functions on the average
performance.

Other than the papers cited above, our work is also related to the
work of Vishwanath \emph{et al.} \cite{Vishwanath} which builds on
\cite{Tse} and takes a similar approach to the rate and power
allocation problem for linear utility functions. Other works address
different criteria for resource allocation including minimizing the
weighted sum of transmission powers \cite{power_min}, and
considering Quality of Service (QoS) constraints \cite{QoS}.

The remainder of this paper is organized as follows: In Section II,
we introduce the model and describe the capacity region of a
multiple-access channel. In Section III, we consider the utility
maximization problem in fading channel, and present a rate
allocation policy and characterize the tracking neighborhood in
terms of the maximum speed of fading. In Section IV, we provide an
alternative rate allocation policy and provide a bound on the size
of tracking neighborhood as a function of the average speed of
fading. Finally, we give our concluding remarks in Section V.

Regarding the notation, we denote by $x_i$ the $i$-th component of a
vector $\bs x$. We denote the nonnegative orthant by
$\mathbb{R}^n_+$, i.e., $\mathbb{R}^n_+ = \{\bs x\in
\mathbb{R}^n\mid \bs x\ge 0\}$. We write $\bs x'$ to denote the
transpose of a vector $\bs x$. The exact projection operation on a
convex set is denoted by $\mathcal P$.

\section{System Model}
We consider $M$ users sharing the same media to communicate to a single receiver. We model the
channel as a Gaussian multiple access channel with flat fading effects
\begin{equation}\label{fading_model}
    Y(n) = \sum_{i=1}^M  \sqrt{H_i(n)} X_i(n) + Z(n),
\end{equation}
where $X_i(n)$ are the transmitted waveform with average power $P_i$, $H_i(n)$ is the channel state
corresponding to the \textit{i}-th user at time slot $n$, and $Z(n)$ is white Gaussian noise with
variance $N_0$. The channel state process is assumed to be ergodic and bounded. We also assume that
the channel states are known to all users and the receiver \footnote{This assumption is satisfied
in practice when the receiver measures the channels and feeds back the channel information to the
users.}. Throughout this work we assume that the transmission powers are fixed and no prior
knowledge of channel statistics is available.

We model the speed of fading as follows:
\begin{equation}\label{fading_speed}
    |H_i(n+1) -  H_i(n) | = V^i_n, \quad \textrm{for all } n, \ i= 1, \ldots, M,
\end{equation}
where $V^i_n$ is a nonnegative random variable bounded from above by $\hat v^i$, and $\{V^i_n\}$
are independent identically distributed (i.i.d.) for fixed $i$. Under slow fading conditions, the
distribution of $V^i_n$ is concentrated around zero.

We first consider the non-fading case where the channel state, $\bs H$, is fixed. The capacity
region of the Gaussian multiple-access channel with no power control is described as follows
\cite{cover}:
\begin{eqnarray}\label{Cg}
    C_g(\bs P, \bs H) &=& \bigg\{ \bs R \in \mathbb{R}^M_+: \sum_{i \in S} R_i \leq  C\Big(\sum_{i \in S} H_i P_i,
    N_0\Big), \nonumber \\
     &&\quad \qquad  \textrm{for all}\  S \subseteq \mathcal M = \{1,\ldots, M\} \bigg\},
\end{eqnarray}
where $P_i$ and $R_i$ are the \emph{i}-th transmitter's power and
rate, respectively. $C(P,N)$ denotes Shannon's formula for the
capacity of additive white Gaussian noise (AWGN) channel given by
\begin{equation}\label{C_AWGN}
    C(P,N) = \frac{1}{2}\log(1+\frac{P}{N}) \quad \textrm{nats}.
\end{equation}

For a multiple-access channel with fading, but fixed transmission powers $P_i$, the
\emph{throughput} capacity region is obtained by averaging the instantaneous capacity regions with
respect to the fading process \cite{Shamai}:
\begin{eqnarray}\label{Ca}
    C_a(\bs P) &=& \bigg\{ \bs R \in \mathbb{R}^M_+: \sum_{i \in S} R_i
    \leq \mathbb{E}_{\bs H} \bigg[ C\Big(\sum_{i \in S} H_i P_i, N_0\Big) \bigg], \nonumber \\
    && \quad \qquad \qquad \qquad  \textrm{for all} \  S  \subseteq \{1, \ldots, M\} \bigg\},
\end{eqnarray}
where $\bs H$ is a random vector with the stationary distribution of the fading process. Let us
define the notion of boundary or dominant face for any of the capacity regions defined above.
\begin{dom_face}\label{dom_face}
The \emph{dominant face} or \emph{boundary} of a capacity region, denoted by $\mathcal{F}(\cdot)$,
is defined as the set of all $M$-tuples in the capacity region such that no component can be
increased without decreasing others while remaining in the capacity region.
\end{dom_face}

\section{Resource Allocation for a Fading Channel}
The goal of the dynamic resource allocation problem is to find a rate
allocation policy, $\mathcal R$, which is a map from the fading state $\bs h$
to the transmission rates, $\mathcal R(\bs h) = (\mathcal{R}_1(\bs h), \ldots,
\mathcal{R}_M(\bs h))$. In the following we define the optimal rate allocation
policy with respect to utility function, $u(\cdot)$.

\begin{optimal_policy}\label{optimal_policy}
[Optimal Policy] The optimal rate
allocation policy denoted by $\mathcal{R}^*(\cdot)$ is a mapping that satisfies
$\mathcal{R}^*(\bs H) \in
C_g\big(\mathcal{P}^*(\bs H),\bs H\big)$ for all $\bs H$, such that
\begin{eqnarray}\label{RAC}
\mathbb{E}_{\boldsymbol{H}} [\mathcal{R}^*(\bs H)] = \bs R^* \in& \textrm{argmax}& \quad u(\bs R)
\nonumber \\
&\textrm{subject to}& \quad \bs R \in  C_a({\bs P})
\end{eqnarray}
\end{optimal_policy}

The utility function $u(\bs R)$ is assumed to satisfy the following
conditions.
\begin{assumption}\label{assumption_u} \emph{The following conditions hold:
\begin{itemize}
\item[(a)] The utility function $u(\bs R)$ is concave with respect to vector $\bs R$.
\item[(b)] $u(\bs R)$ is monotonically non-decreasing with respect to $R_i$, for $i = 1, \ldots , M$.
\item[(c)] There exists a scalar $B$ such that
$$\|\bs g \| \leq B, \quad \textrm{for all}\ \bs g \in \partial u(\bs R) \hbox{ and all } \bs R,$$
where $\partial u(\bs R)$ denotes the subdifferential of $u$ at $\bs R$, i.e., the set of all
subgradients \footnote{The vector $\bs g$ is a subgradient of a concave function $f:D \rightarrow \mathbb{R}$ at $x_0$,
if and only if $f(x) - f(x_0) \leq \bs g'(x-x_0)$ for all $x \in D$.} of $u$ at $\bs R$.
\item[(d)] If $\bs R^\dag = \textrm{argmax}_{\bs R \in C_g(\bs P, \bs H)} u(\bs R)$, then there exists a positive
scalar $A$ such that
$$|u(\bs R^\dag) - u(\bs R)| \geq A \|\bs R^\dag - \bs R\|^2, \quad \textrm{for all } \bs R \in C_g(\bs P, \bs H).$$
\end{itemize}
}\label{utility_assump}
\end{assumption}

Assumption \ref{assumption_u}(c) imposes a bound on subgradients of the utility function. In this
paper, it is sufficient to have a utility function with bounded subgradient only in a neighborhood
of optimal solution, but weakening Assumption \ref{assumption_u}(c) may require unnecessary
technical details. Assumption \ref{assumption_u}(d) is a strong concavity type assumption which is
satisfied for most of the utility functions. In fact, strong concavity of the utility implies
Assumption \ref{assumption_u}(d), but it is not necessary.

\begin{greedy_policy}\label{greedy_policy}
[Greedy Policy] A \emph{greedy} rate allocation policy, denoted by
$\mathcal{\bar{R}}$, is given by
\begin{eqnarray}\label{RAC_greedy}
\mathcal{\bar{R}}(\bs H) = & \textrm{argmax}& \quad u(\bs R)
\nonumber \\
&\textrm{subject to}& \quad \bs R \in  C_g({\bs P}, \bs H)
\end{eqnarray}
i.e., for each channel state, the greedy policy chooses the rate vector that maximizes the utility
function over the corresponding capacity region.
\end{greedy_policy}

 Note that the greedy policy is not necessarily optimal for
general concave utility functions, i.e., the expected achieved rate does not maximize the utility
over the throughput capacity region. However, the performance difference, i.e., utility difference
between the expected rates assigned by the greedy and the optimal policy, is bounded and the bounds
can be characterized in terms of channel variations and the structure of the utility function
\cite{MAC_Wiopt}.

The maximization problem in (\ref{RAC_greedy}) is a convex program
and the optimal solution can be obtained by iterative methods such
as the gradient projection method with approximate projection
studied in \cite{MAC_asilomar}. The $k$-th iteration of this method
is given by
    \begin{equation}\label{iteraion}
        \bs R^{k+1} = \tilde{\mathcal P}(\bs R^k + \alpha ^k \bs g^k), \quad \bs g^k \in \partial
        u(\bs R^k),
    \end{equation}
    where $\bs g^k$ is a subgradient of $u$ at $\bs R^k$, and $\alpha ^k$ denotes
    the stepsize. $\tilde{\mathcal P}$ denotes the approximate projection operator which is defined
    in the following.

\begin{approx_proj}\label{approx_proj_def}
    Let $X = \{\bs x \in \mathbb{R}^n| A\bs x \leq \bs b\}$ where $A$ has non-negative entries. Let $\bs y \in
    \mathbb{R}^n$ violate the constraint $\bs a_i' \bs x \leq b_i$, for $i\in\{i_1, \ldots, i_l\}$. The approximate
    projection of $\bs y$ on $X$, denoted by $\tilde{\mathcal P}$, is given by
    $$ \tilde{\mathcal P}(\bs y) = \mathcal P_{i_1}(\ldots (\mathcal P_{i_{l-1}}(\mathcal P_{i_l}(\bs y)))),$$
    where $\mathcal P_{i_k}$ denotes the exact projection on the hyperplane $\{\bs x \in \mathbb{R}^n| \bs a_{i_k}' \bs x =
    b_{i_k}\}$.
\end{approx_proj}

\begin{figure}
  \centering
  \includegraphics[width=.3\textwidth]{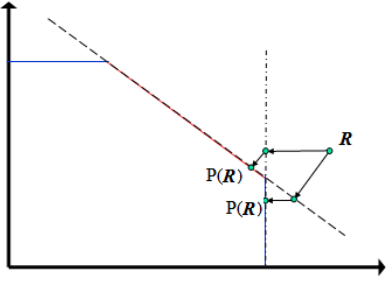}\\
  \caption{Approximate projection of $\bs R$ on a two-user MAC region}\label{projection_fig}
\end{figure}

An example of approximate projection on a two-user multiple-access capacity region is illustrated
in Figure \ref{projection_fig}. Note that the result of projection is not necessarily unique.
However it is pseudo-nonexpansive, i.e., the distance between any feasible point and the projected
point is smaller than its distance to the original point. Under Assumption \ref{assumption_u} and
specific stepsize rules, we established the convergence of the iterations in (\ref{iteraion}) to
the optimal solution in (\ref{RAC_greedy}) by using the pseudo-nonexpansiveness of the approximate
projection (see \cite{MAC_asilomar}, Proposition 2). We also showed by exploiting the polymatroid
structure of the capacity region, each iteration in (\ref{iteraion}) can be computed in $O(M^3\log
M )$ time. However, for each channel state, finding even a "near-optimal" solution of the problem
in (\ref{RAC_greedy}) requires a large number of iterations, making the online evaluation of the
greedy policy impractical. In the following section, we introduce an alternative rate allocation
policy, which implements a single gradient projection iteration of the form (\ref{iteraion}) per
time slot.

\section{Approximate Rate Allocation Policy}
In this section, we assume that the channel state information is available instantly at each time
slot $n$, and the computational resources are limited such that a single iteration of the gradient
projection method in (\ref{iteraion}) can be implemented in each time slot.


\begin{approximate_policy}\label{approximate_policy}
[Approximate Policy] Given some fixed integer $k \geq 1$, we define the \emph{approximate} rate allocation policy,
$\widetilde{\mathcal{R}}$,
as follows:
$$ \widetilde{\mathcal{R}}\big(\bs H(0)\big) = \bar{\mathcal{R}}\big(\bs H(0)\big) ,$$
 \begin{equation}\label{policy_worst}
     \widetilde{\mathcal{R}}\big(\bs H(n)\big) = \bs{R}_t^\tau, \quad \textrm {for all } n \geq 1,
 \end{equation}
 where
\begin{equation}\label{tau_max}
    \tau = \textrm{arg}\!\!\!\!\!\!\max_{0 \leq j < k-1} \ u(\bs{R}^j_t), \quad t = \left\lfloor \frac{n-1}{k} \right\rfloor,
\end{equation}
and $\bs R^j_t \in \mathbb R^M$ is given by the following gradient projection iterations
\begin{eqnarray}\label{iteration_tv}
    \bs R_t^0 &=& \tilde P_{t}\left[ \widetilde{\mathcal{R}}\big(\bs H(kt)\big)\right],  \nonumber \\
    \bs R_t^{j+1} &=& \tilde P_{t}\left[ \bs R_t^j + \alpha^j \bs g^j
    \right], \quad j=1,\ldots,k-1,
\end{eqnarray}
where $\bs g^j$ is a subgradient of $u(\cdot)$ at $\bs R^j$, $\alpha^j$ denotes the stepsize and
$\tilde P_{t}$ is the approximate projection on $C_g(\bs P, \bs H(kt))$.
\end{approximate_policy}

For $k=1$, (\ref{iteration_tv}) reduces to taking only one gradient projection iteration at each
time slot. For $k >1$, the proposed rate allocation policy essentially let the channel state change
for a block of $k$ consecutive time slots, and then takes $k$ iterations of the gradient projection
method with approximate projection. Note that to compute the policy at time slot $n$, we are using
the channel state information at time slots $kt, k(t-1), \ldots$ Hence, in practice the channel
measurements need to be computed every $k$ time slots.

 There is a tradeoff in choosing $k$, because taking only
one gradient projection step may not be sufficient to get close enough to the greedy policy's
operating point. Moreover, for large $k$ the new operating point of the greedy policy can be far
from the previous one, and $k$ iterations may be insufficient again. So the parameter $k$ should be
chosen optimally to obtain the best performance for the approximate policy.

Before stating the main result, let us introduce some definitions and lemmas.

       \begin{expansion}\label{expansion_def}
        Let $Q$ be a polyhedron described by a set of linear constraints, i.e.,
        \begin{equation}\label{polyhedron}
            Q = \left\{\bs x \in \mathbb{R}^n: A \bs x \leq \bs b \right\}.
        \end{equation}
        Define the \emph{expansion} of $Q$ by $\delta$, denoted by $\mathcal{E}_\delta(Q)$, as the polyhedron
        obtained by relaxing all the constraints in (\ref{polyhedron}), i.e., $ \mathcal{E}_\delta(Q) = \left\{\bs x \in \mathbb{R}^n: A \bs x \leq \bs b + \delta\mathbf{1}
            \right\},$
        where $\mathbf{1}$ is the vector of all ones.
        \end{expansion}

        \begin{Hausdorff}\label{Hausdorff_def}
        Let $X$ and $Y$ be two polyhedra described by a set of linear constraints. Let
        $\mathcal{E}_d(X)$ be an \emph{expansion} of $X$ by relaxing its constraints by $d$. The distance
        $d_H(X,Y)$ between $X$ and $Y$ is defined as the minimum scalar $d$ such that $X \subseteq \mathcal{E}_d(Y)$ and $ Y \subseteq
        \mathcal{E}_d(X)$.
        \end{Hausdorff}

The next lemma shows that if the distance between two capacity regions is small, the distance
between the optimal solutions of maximizing the utility function over these regions is also small.

\begin{opt_dist}\label{opt_dist}
            Let $\bs H_1$ and $\bs H_2$ be two different channel states. Also, let $\bs R_1^*$ and
            $\bs R_2^*$ be the optimal solution of maximizing a utility function over $C_g(\bs P, \bs H_1)$ and $C_g(\bs P, \bs H_2)$,
            respectively. If the utility satisfies Assumption \ref{assumption_u}, and
            $$d_H\big(C_g(\bs P, \bs H_1), C_g(\bs P, \bs H_2)\big) \leq \delta$$
            Then, we have
            \begin{equation}\label{opt_dist_result}
                \|\bs R_1^* - \bs R_2^*\| \leq
                {\delta}^{\frac{1}{2}}\left[{\delta}^{\frac{1}{2}}+\Big(\frac{B}{A}\Big)^{\frac{1}{2}}\right].
            \end{equation}
\end{opt_dist}

            \begin{proof}
            See Appendix \ref{appendix}.
            \end{proof}

In the following lemma, we translate the model for the speed of fading in terms of channel state
variations into changes in the corresponding capacity regions.

\begin{region_dist}\label{region_dist}
Let $\{[H_i(n)]_{i=1,\ldots,M}\}$ be the fading process that satisfies condition in
(\ref{fading_speed}). We have
\begin{equation}\label{capacity_dist}
    d_H\Big(C_g\big(\bs P, \bs H(n+1)\big),C_g\big(\bs P, \bs H(n)\big)\Big) \leq W_n,
\end{equation}
where $\{W_n\}$ are nonnegative independent identically distributed random variables bounded from
above by $\hat w = \frac{1}{2} \sum_{i=1}^M \hat v^i P_i$, where $\hat v^i$ is an upperbound on the
process $\{V_n^i\}$ and $P_i$ is the $i$-th user's transmission power.
\end{region_dist}
\begin{proof}
By Definition \ref{Hausdorff_def} we have
\begin{eqnarray}\label{region_lemma1}
    &&d_H\Big(C_g\big(\bs P, \bs H(n+1)\big),C_g\big(\bs P, \bs H(n)\big)\Big)  \nonumber \\
    &&= \max_{S \subseteq \mathcal M} \frac{1}{2} \bigg |\log\Big( 1+ \frac{\sum_{i \in S}(H_i(n+1)-H_i(n))P_i}{1+ \sum_{i \in S} H_i(n)P_i}
    \Big)\bigg|  \nonumber \\
    && \leq \max_{S \subseteq \mathcal M} \frac{\sum_{i \in S}|H_i(n+1)-H_i(n)|P_i}{2(1+ \sum_{i \in S}
    H_i(n)P_i)} \nonumber \\
    && \leq \frac{1}{2} \sum_{i=1}^M|H_i(n+1)-H_i(n)|P_i =\frac{1}{2} \sum_{i=1}^M V^i_n P_i.
\end{eqnarray}
Therefore, (\ref{capacity_dist}) is true for $W_n = \frac{1}{2} \sum_{i=1}^M V^i_n P_i$. Since
$V^i_n$'s are i.i.d. and bounded above by $\hat v^i_n$, $W_n$'s are i.i.d. and bounded from above
by $\frac{1}{2} \sum_{i=1}^M \hat v^i P_i$.

\end{proof}
The following lemma by Nedi\'c and Bertsekas \cite{subgradient_rate}
addresses the convergence rate of the gradient projection method
with constant stepsize.

\begin{subgradient_rate}\label{subgradient_rate}
Let rate allocation policies $\bar{\mathcal{R}}$  and $\widetilde{\mathcal{R}}$ be given by
Definition \ref{greedy_policy} and Definition \ref{approximate_policy}, respectively. Also, let Assumption \ref{assumption_u}
hold and the stepsize $\alpha^n$ be fixed to some positive constant $\alpha$. Then for a positive
scalar $\epsilon$ we have
\begin{equation}\label{greedy_worst_rate}
    u\Big(\widetilde{\mathcal{R}}\big(\bs H(n)\big)\Big) \geq  u\Big(\bar{\mathcal{R}}\big(\bs
    H(kt)\big)\Big) - \frac{\alpha B^2 + \epsilon}{2},
\end{equation}
where $k$ satisfies
\begin{equation}\label{k_condition}
    k \geq \bigg \lfloor \frac{\|\bs R_t^{0} - \bar{\mathcal{R}}\big(\bs
    H(kt)\big) \|^2}{\alpha \epsilon} \bigg \rfloor.
\end{equation}

\end{subgradient_rate}

\begin{proof}
See Proposition 2.3 of \cite{subgradient_rate}.
\end{proof}

We next state our main result, which shows that the approximate rate allocation
policy given by Definition \ref{approximate_policy} tracks the greedy policy
within a neighborhood which is quantified as a function of the maximum speed of
fading, the parameters of the utility function, and the transmission
powers.

\begin{tracking_worst}\label{tracking_worst}
Let Assumption \ref{assumption_u} hold and the rate allocation policies  $\bar{\mathcal{R}}$  and
$\widetilde{\mathcal{R}}$ be given by Definition \ref{greedy_policy} and Definition
\ref{approximate_policy}, respectively. Let $k = \lfloor(\frac{2B}{A w'})^{\frac{2}{3}}\rfloor$ and
fix the stepsize to $\alpha = \big(\frac{16A}{B^2}\big)^\frac{1}{3} {w'}^\frac{2}{3}$ in Eq.\
(\ref{iteration_tv}), where $w' = {\hat w}^{\frac{1}{2}}\big(\hat
w^{\frac{1}{2}}+(\frac{B}{A})^{\frac{1}{2}}\big)$, $\hat w$ is the upperbound on $W_n$ as defined
in Lemma \ref{region_dist}, $A$ and $B$ are system parameters depending on the structure of utility
function as in Assumption \ref{assumption_u}(c),(d). Then, we
have
\begin{equation}\label{greedy_worst_dist}
    \|\widetilde{\mathcal{R}}\big(\bs H(n)\big) - \bar{\mathcal{R}}\big(\bs H(n)\big)\| \leq
    2\theta = 2(\frac{2B}{A})^{\frac{2}{3}} w'^{\frac{1}{3}}.
\end{equation}

\end{tracking_worst}
\begin{proof}
First, we show that
\begin{equation}\label{tracking1}
    \|\widetilde{\mathcal{R}}\big(\bs H(n)\big) - \bar{\mathcal{R}}\big(\bs H(kt)\big)\| \leq
    \theta = (\frac{2B}{A})^{\frac{2}{3}} w'^{\frac{1}{3}},
\end{equation}
where $t = \lfloor \frac{n-1}{k} \rfloor$. The proof is by induction on $t$. For $t=0$ the claim is
trivially true. Now suppose that (\ref{tracking1}) is true for some positive $t$. Hence, it also
holds for $n = k(t+1)$ by induction hypothesis, i.e.,
\begin{equation}\label{tracking2}
   \|\bs R_{t+1}^{0} - \bar{\mathcal{R}}\big(\bs H(kt)\big) \| \leq \theta.
\end{equation}
On the other hand, by Lemma \ref{opt_dist} and Lemma \ref{region_dist} we have
\begin{equation}\label{tracking3}
   \|\bar{\mathcal{R}}\big(\bs H(k(t+1))\big) - \bar{\mathcal{R}}\big(\bs H(kt)\big) \| \leq kw'
   \leq \theta.
\end{equation}
Therefore, by triangle inequality we have the following
\begin{equation}\label{tracking4}
       \|\bs R_{t+1}^{0} - \bar{\mathcal{R}}\big(\bs H(k(t+1))\big) \| \leq 2\theta.
\end{equation}
After plugging the corresponding values of $\alpha$ and $\theta$, it is straightforward to show
that (\ref{k_condition}) holds for $\epsilon = \alpha B^2$. Thus, we can apply Lemma
\ref{subgradient_rate} to show
\begin{equation}\label{tracking5}
    \bigg|u\Big(\widetilde{\mathcal{R}}\big(\bs H(n)\big)\Big) - u\Big(\bar{\mathcal{R}}\big(\bs
    H(k(t+1))\big)\Big)\bigg| \leq \alpha B^2.
\end{equation}

By Assumption \ref{assumption_u}(d) we can write
\begin{equation}\label{tracking6}
    \|\widetilde{\mathcal{R}}\big(\bs H(n)\big) - \bar{\mathcal{R}}\big(\bs H(k(t+1))\big)\| \leq
    \big(\frac{\alpha B^2}{A}\big)^\frac{1}{2} = \theta.
\end{equation}
Therefore, the proof of (\ref{tracking1}) is complete by induction.

Again by applying Lemma \ref{opt_dist} and Lemma \ref{region_dist} we have
\begin{equation}\label{tracking7}
   \|\bar{\mathcal{R}}\big(\bs H(n)\big) - \bar{\mathcal{R}}\big(\bs H(kt)\big) \| \leq kw'
   \leq \theta,
\end{equation}
and the desired result directly follows from (\ref{tracking1}) and (\ref{tracking7}) by triangle
inequality.
\end{proof}

It is straightforward to show that the parameters $k$ and $\alpha$ in Theorem \ref{tracking_worst}
are designed such that the smallest tracking neighborhood, $\theta$, is obtained for the approximate policy
presented in Definition \ref{approximate_policy} with constant stepsize. The proof is by parameterizing $\theta$,
the size of the tracking neighborhood, in terms of $k$ and minimizing $\theta(k)$ by relaxing $k$
to be a real and differentiating with respect to $k$. We eliminate the full proof for brevity.
Theorem \ref{tracking_worst} provides a bound on the size of the tracking neighborhood as a
function of the maximum speed of fading, denoted by $\hat w$, which may be too conservative. It is
of interest to provide a rate allocation policy and a bound on the size of its tracking
neighborhood as a function of the average speed of fading. The next section addresses this issue.

%
%
%

\section{Improved Approximate Rate Allocation Policy}
In this section, we design an efficient rate allocation policy that tracks the greedy policy within
a neighborhood characterized by the average speed of fading which is typically much smaller than
the maximum speed of fading. We consider policies which can implement one gradient projection
iteration per time slot.

Unlike the approximate policy given by (\ref{policy_worst}) which uses the channel state information once in
every $k$ time slots, we present an algorithm which uses the channel state information in all time
slots. Roughly speaking, this method takes fixed a number of gradient projection iterations only after the
change in the channel state has reached a certain threshold.

\begin{improved_approximate_policy}\label{improved_approximate_policy}
[Improved Approximate Policy] Let $\{W_n\}$ be the  sequence of nonnegative random variables as defined in Lemma
\ref{region_dist}, and $\gamma$ be a positive constant. Define the sequence $\{T_i\}$ as
\begin{eqnarray}\label{T_def}
  T_0 &=& 0, \nonumber \\
  T_{i+1} &=& \min\{ t \mid \sum_{n = T_i}^{t-1} W_n \geq \gamma \}.
\end{eqnarray}

Define the \emph{improved approximate} rate allocation policy, $\widehat{\mathcal{R}}$, with parameters $\gamma$ and $k$, as
the following:
$$ \widehat{\mathcal{R}}\big(\bs H(0)\big) = \bar{\mathcal{R}}\big(\bs H(0)\big) ,$$
 \begin{equation}\label{policy_avg}
     \widehat{\mathcal{R}}\big(\bs H(n)\big) = \bs{R}_t^\tau, \quad \textrm {for all } n \geq 1,
 \end{equation}
 where
\begin{eqnarray}\label{tau_avg}
    t &=& \max\{i \mid T_i < n\}, \\
    \tau &=& \textrm{arg}\!\!\!\!\!\max_{0 \leq j < k-1} \ u(\bs{R}^j_t),
\end{eqnarray}

and $\bs R^j_t \in \mathbb R^M$ is given by the following gradient projection iterations
\begin{eqnarray}\label{iteration_tv_avg}
    \bs R_t^0 &=& \tilde P_{t}\left[ \widehat{\mathcal{R}}\big(\bs H(T_t)\big)\right],  \nonumber \\
    \bs R_t^{j+1} &=& \tilde P_{t}\left[ \bs R_t^j + \alpha^j \bs g^j
    \right], \quad j=1,\ldots,k-1,
\end{eqnarray}
where $\bs g^j$ is a subgradient of $u(\cdot)$ at $\bs R^j$, $\alpha^j$ denotes the stepsize and
$\tilde P_{t}$ is the approximate projection on $C_g(\bs P, \bs H(T_t))$.
\end{improved_approximate_policy}

\begin{iteration_queue}\label{iteration_queue}
Let $t$ be as defined in (\ref{tau_avg}), and let $\bar w$ denote the expected value of $W_n$. If
$k = \frac{\gamma}{\bar w}$, then we have
\begin{equation}\label{almost_sure_nk}
    \lim_{n \rightarrow \infty} \frac{n}{tk} = 1, \quad \textrm{with probability } 1.
\end{equation}

\end{iteration_queue}

\begin{proof}
The sequence $\{T_i\}$ is obtained as the random walk generated by $W_n$'s cross the threshold
level $\gamma$. Since $W_n$'s are positive random variables, we can think of the threshold crossing
as a renewal process, denoted by $N(\cdot)$, with inter arrivals $W_n$.

We can rewrite the limit as follows
\begin{equation}\label{queue1}
     \lim_{n \rightarrow \infty} \frac{n-N(t\gamma) + N(t\gamma)}{tk} =  \lim_{n \rightarrow \infty}
     \frac{n-N(t\gamma)}{tk} + \bar w \frac{N(t\gamma)}{t\gamma}.
\end{equation}
Since the random walk will hit the threshold with probability 1, the first term goes to zero with
probability 1. Also, by Strong law for renewal processes the second terms goes to 1 with
probability 1 (see \cite{dsp}, p.60).
\end{proof}
Theorem \ref{iteration_queue} essentially guarantees that the number of gradient projection
iterations is the same as the number of channel measurements in the long run with probability 1.

\begin{tracking_avg}\label{tracking_avg}
Let Assumption \ref{assumption_u} hold and the rate allocation policies  $\bar{\mathcal{R}}$  and
$\widehat{\mathcal{R}}$ be given by
Definition \ref{greedy_policy} and Definition \ref{improved_approximate_policy}, respectively.
Also, let $\gamma = c(\frac{B}{A})^\frac{3}{4} \bar w^\frac{1}{4}$, and $k = \lfloor
\frac{\gamma}{\bar w}\rfloor$ and fix the stepsize to $\alpha = \frac{A\gamma^2}{B^2}$ in
(\ref{iteration_tv_avg}), where $c \geq 1$ is a constant satisfying the following equation
\begin{equation}\label{c_equation}
    \frac{(c^2-1)^8}{2^8 c^4} = \hat w.
\end{equation}
Then
\begin{equation}\label{tracking_avg_dist}
    \|\widehat{\mathcal{R}}\big(\bs H(n)\big) - \bar{\mathcal{R}}\big(\bs H(n)\big)\| \leq
    2\gamma + (\frac{\gamma B}{A})^\frac{1}{2}.
\end{equation}

\end{tracking_avg}

\begin{proof}
We follow the line of proof of Theorem \ref{tracking_worst}. First, by induction on $t$ we show
that
\begin{equation}\label{tracking_avg1}
 \|\widehat{\mathcal{R}}\big(\bs H(n)\big) - \bar{\mathcal{R}}\big(\bs H(T_t)\big)\| \leq
    \gamma,
\end{equation}
where $t$ is defined in (\ref{tau_avg}). The base is trivial. Similar to (\ref{tracking2}), by
induction hypothesis we have
\begin{equation}\label{tracking_avg2}
   \|\bs R_{t+1}^{0} - \bar{\mathcal{R}}\big(\bs H(T_t)\big) \| \leq \gamma.
\end{equation}
By definition of $T_i$ in (\ref{T_def}) we can write
\begin{equation}\label{tracking_avg3}
    d_H\Big(C_g\big(\bs P, \bs H(T_{t+1})\big),C_g\big(\bs P, \bs H(T_t)\big)\Big) \leq \gamma.
\end{equation}
Thus, by Lemma \ref{opt_dist}, we have
\begin{equation}\label{tracking_avg4}
     \|\bar{\mathcal{R}}\big(\bs H(T_{t+1})\big) - \bar{\mathcal{R}}\big(\bs H(T_t)\big) \| \leq
     \gamma^\frac{1}{2}\big(\gamma^\frac{1}{2} + (\frac{B}{A})^\frac{1}{2}\big).
\end{equation}
Therefore, by combining (\ref{tracking_avg2}) and (\ref{tracking_avg4}) by triangle inequality we
obtain
\begin{equation}\label{tracking_avg5}
   \|\bs R_{t+1}^{0} - \bar{\mathcal{R}}\big(\bs H(T_{t+1})\big) \| \leq 2\gamma + (\frac{\gamma
   B}{A})^\frac{1}{2}.
\end{equation}
Using the fact that $\bar w \leq \hat w = \frac{(c^2-1)^8}{2^8 c^4}$, after a few steps of
straightforward manipulations we can show that
\begin{equation}\label{tracking_avg6}
    \|\bs R_{t+1}^{0} - \bar{\mathcal{R}}\big(\bs H(T_{t+1})\big) \|^2 \leq \Big(2\gamma + (\frac{\gamma   B}{A})^\frac{1}{2}\Big)^2 \leq c^4 \frac{\gamma
    B}{A}.
\end{equation}
Now by plugging in (\ref{k_condition}) the values of $\alpha$ and $\gamma$ in terms of system
parameters we can verify that
\begin{equation}\label{tracking_avg7}
    k = \left\lfloor \frac{\gamma}{\bar w}\right\rfloor =  \bigg\lfloor \frac{c^4\frac{\gamma B}{A}}{A\frac{\gamma^2}{B^2} A\gamma^2}
    \bigg\rfloor \geq \bigg\lfloor \frac{\|\bs R_{t+1}^{0} - \bar{\mathcal{R}}\big(\bs H(T_{t+1})\big) \|^2}{\alpha \epsilon}
    \bigg\rfloor.
\end{equation}
Hence, we can apply Lemma \ref{subgradient_rate} for $\epsilon = A\gamma^2$, and conclude
\begin{equation}\label{tracking_avg8}
    \bigg|u\Big(\widehat{\mathcal{R}}\big(\bs H(n)\big)\Big) - u\Big(\bar{\mathcal{R}}\big(\bs
    H(T_{t+1})\big)\Big)\bigg| \leq \alpha B^2.
\end{equation}
By exploiting Assumption \ref{assumption_u}(d) we have
\begin{equation}\label{tracking_avg9}
    \|\widehat{\mathcal{R}}\big(\bs H(n)\big) - \bar{\mathcal{R}}\big(\bs H(T_{t+1})\big)\| \leq
    \big(\frac{\alpha B^2}{A}\big)^\frac{1}{2} = \gamma.
\end{equation}
Therefore, the proof of (\ref{tracking_avg1}) is complete by induction. Similarly to
(\ref{tracking_avg4}) we have
\begin{equation}\label{tracking_avg10}
     \|\bar{\mathcal{R}}\big(\bs H(n)\big) - \bar{\mathcal{R}}\big(\bs H(T_t)\big) \| \leq
     \gamma^\frac{1}{2}\big(\gamma^\frac{1}{2} + (\frac{B}{A})^\frac{1}{2}\big),
\end{equation}
and (\ref{tracking_avg_dist}) follows immediately from (\ref{tracking_avg1}) and
(\ref{tracking_avg10}) by invoking triangle inequality.
\end{proof}

Theorem \ref{iteration_queue} and Theorem \ref{tracking_avg}
guarantee that the presented rate allocation policy tracks the
greedy policy within a small neighborhood while with probability 1,
only one gradient projection iteration is computed per time slot.
The neighborhood is characterized in terms of the average behavior
of channel variations and vanishes as the fading speed decreases.

\section{Conclusion}
We study the problem of rate allocation in a fading multiple access
channel with no power control from an information theoretic point of
view. Our goal is to approximate the optimal rate allocation policy,
which yields an average rate that maximizes a general concave
utility function of transmission rates over the throughput capacity
region of the multiple-access channel.

We present a dynamic rate allocation policy which takes a block of
channel measurements and implements the same number of gradient
projection iterations with approximate projection at the end of each
block. This rate allocation policy tracks the greedy policy within a
small neighborhood whose size decreases as a function of the maximum
speed of fading.

In order to provide a bound on the tracking neighborhood in terms of
average speed of fading, we present an alternative rate allocation
policy. This policy adaptively selects variable block lengths for
channel measurements using feedback information about the current
channel states. It implements a fixed number of gradient iterations
at the end of these blocks. We show that the ratio of the total
number of channel measurements and the number of gradient iterations
converges to 1 with probability one. We also provide a bound on the
size of the neighborhood with which the new policy tracks the greedy
policy as a function of the average speed of fading. The proposed
dynamic rate allocation policies are efficiently implementable since
they require a single gradient projection step per channel
measurement and the projection can be done in time polynomial in the
number of users.


\bibliographystyle{unsrt}
\bibliography{MAC}

\appendices
\section{Proof of Lemma \ref{opt_dist} }\label{appendix}

                Without loss of generality assume that $u(\bs R_2^*) \geq u(\bs R_1^*)$. To
                simplify the notations for capacity regions, let $C_1 = C_g\big(\bs P, \bs H_1\big)$ be a \emph{polymatroid}, i.e.,
            \begin{equation}\label{polymatroid}
                C_1 = \bigg\{ \bs R \in \mathbb{R}^M_+: \sum_{i \in S} R_i \leq f(S),\
                \textrm{for all}\ S \subseteq \mathcal M \bigg\},
            \end{equation}
             for some submodular function $f(S)$, and $C_2$ be an \emph{expansion} of $C_1$ by
            $\delta$. We first show that for every $\bs R \in \mathcal{F}(C_2)$, there exists a vector
                $\bs R' \in \mathcal{F}(C_1)$ such that $\|\bs R - \bs R'\| \leq \delta$, where $\mathcal
                F(\cdot)$ denotes the dominant face of a capacity region as in Definition \ref{dom_face}.

                Assume $R$ is a vertex of $C_2$. Then the polymatroid structure of $C_2$ implies
                that $R$ is the intersection of $M$ constraints corresponding to a chain of subsets
                of $\mathcal{M}$. Hence, there is some $k \in \mathcal{M}$ such that $ R_k = f(\{k\}) + \delta
                $. Choose $\bs R'$ as follows

                 \begin{equation}\label{R'_R}
                   R'_i = \left\{ \begin{array}{ll}
                   R_i - \delta, & \textrm{$i = k$}\\
                   R_i, & \textrm{otherwise.}
                   \end{array} \right.
                 \end{equation}

                $\bs R'$ is obviously in a $\delta$-neighborhood of $\bs R$. Moreover, the
                constraint corresponding to the set $\mathcal{M}$ is active for $\bs R'$, so we
                just need to show that $R'$ is feasible in order to prove that it is on the
                dominant face. First, let us consider the sets $S$ that contain $k$. We have
                \begin{equation}\label{k_in_S}
                    \sum_{i \in S} R'_i = \sum_{i \in S} R_i - \delta \leq f(S).
                \end{equation}
                Second, consider the case that $k \notin S$.
                \begin{eqnarray}
                  \sum_{i \in S} R'_i &=& \sum_{i \in S \cup \{k\}} R'_i - R_k + \delta \nonumber \\
                   &\leq& f(S \cup \{k\})  + \delta - R_k \nonumber \\
                   &\leq& f(S) + f(\{k\}) + \delta - R_l \nonumber \\
                   &=& f(S). \nonumber
                \end{eqnarray}

                where the first inequality come from (\ref{k_in_S}), and the second inequality is
                valid because of the submodularity of the function $f(\cdot)$.

               The previous argument establishes that the claim is true for each vertex $\bs R_j$ of the dominant face. But
                every other point $\bs R$ on the dominant face can be represented as
                a convex combination of the vertices, i.e.,
                $$ \bs R = \sum_j \alpha_j \bs R_j, \qquad  \sum_j \alpha_j = 1, \alpha_j \geq 0.$$
                 Using the convexity of the norm function, it is quite straightforward to show that the desired $\bs R'$
                 is given by
                $$ \bs R' = \sum_j \alpha_j \bs R'_j,$$
                where $\bs R'_j$ is obtained for each $\bs R_j$ in the same manner as in
                (\ref{R'_R}).

                 So we have shown that there exists some $\bs R$ on the dominant face of $C_1 = C_g(\bs P, \bs H_1)$ such
                 that $\|\bs R_2^* - \bs R\| \leq \delta$. Thus, from the hypothesis and the fact that $u(\bs R_2^*) \geq u(\bs R_1^*) \geq u(\bs R)$, we have
                \begin{equation}\label{opt_dist1}
                    u(\bs R_2^*) - u(\bs R) = |u(\bs R_2^*) - u(\bs R)| \leq B\|\bs R_2^* - \bs R\| \leq B
                    \delta.
                \end{equation}
                Now suppose that $\|\bs R_1^* - \bs R\| > (\frac{B}{A} \delta)^{\frac{1}{2}}$, hence
                by Assumption \ref{assumption_u}(d) we have
                \begin{equation}\label{opt_dist2}
                    u(\bs R_1^*) - u(\bs R) = |u(\bs R_1^*) - u(\bs R)| >  B \delta.
                \end{equation}
                By subtracting (\ref{opt_dist1}) from (\ref{opt_dist2}) we obtain $u(\bs R_2^*) <
                u(\bs R_1^*)$ which is a contradiction. Therefore, $\|\bs R_1^* - \bs R\| \leq
                (\frac{B}{A} \delta)^{\frac{1}{2}}$, and the desired result follows immediately by invoking the
                triangle inequality.

\end{document}